\documentclass[twocolumn,prl,superscriptaddress,floatfix,showpacs]{revtex4-1}
\usepackage[utf8]{inputenc}
\usepackage{amsmath}
\usepackage{amssymb}
\usepackage{graphicx}
\usepackage{color}

\makeatletter
\usepackage{graphics}
\usepackage{epsfig}
\usepackage{color}

\newcommand{\be}{\begin{equation}} \newcommand{\ee}{\end{equation}}
\newcommand{\bea}{\begin{eqnarray}} \newcommand{\eea}{\end{eqnarray}}

\makeatother

\begin{document}

\title{Exact probability distribution function for the volatility of cumulative production}

\author{Rubina Zadourian}

\affiliation{Institute for New Economic Thinking at the Oxford Martin School, Oxford, UK}

\author{Andreas Kl\"umper}

\affiliation{Wuppertal University, D-42119 Wuppertal, Germany}

\date{\today}

\begin{abstract}
In this paper we study the volatility and its probability distribution
function for the cumulative production based on the experience curve
hypothesis. This work presents a generalization of the study of volatility in
\cite{Lafond}, which addressed the effects of normally distributed noise in the production process. Due to its wide applicability in industrial and technological
activities we present here the mathematical foundation for an arbitrary distribution function of the process, which we expect will pave the future research on
production and market strategy.
\end{abstract}
\maketitle
\section*{Introduction}

Understanding the volatile behaviour of industrial activities and the
complexity related to them, intrigues many researchers. One of the stylized
facts for describing this phenomenon is a well known experience
curve. The concept of experience curves and empirical evidence
  for them were presented in Wright’s \cite{Wright} seminal paper, in which
he first discovered the relationship of cost and quantity. Wright’s curve is
known in the literature as ``learning curve'', as it is based on ``the more
learning by more producing hypothesis'', for describing the price-experience
relationship. Wright realized that empirically the reduction of cost followed
a constant proportion rate, as the production duplicated. In other words, the
higher the experience in producing a specific product is, the lower its costs
are, when the inflation is factored out.

%%%Furthermore this hypothesis sheds light to a new relationship for risk
%%%management and portfolio optimization sector, by considering a stability in
%%%equilibrium systems.

The inspiration of this paper comes from the fact that the experience curves
hypothesis can provide a significant understanding of the market strategy, for
instance export potentials due to the knowledge of experience levels, the
prediction of future prices, given some information about the market costs
decrease by some consistent rate of decline, the applicability in risks
management, etc. (Note that the notion is suitable for cost control or
forecasting over long range strategic development).

The phenomenon depends on some crucial factors, i.e.~competent management,
technological improvement, etc. Furthermore there must be a characteristic
pattern that causes this phenomenon, for instance a better development of
better tools, automatization, training programs \cite{Terwiesch,Vits, Serel,
  Azizi}, prior experience and the work complexity task
\cite{Nembhard,Pananiswami}.

The notion of experience curve could also describe the effect between business
competitors, for example, who is faster by reducing the costs, which is 
an example of complex systems interactions and network.

There is a vast literature on empirical information about the experience
curves, including a wide range of industrial activities, see
e.g.~\cite{Arrow}. The aforementioned work had a major impact on the development of this concept, by
arguing that technical learning was a result of experience gained, based
on the idea learning by doing. Some researchers question its usefulness for
forecasting and planning the deployment of industrial and technological
activities \cite{Lafond, Ayres, Anzanello, Sahal, Martino}. In the aforementioned
literature it has been found that experience curves can be used to estimate
future technology costs, considering the shape of the forecast error
distribution. (Note the finding depends on some parameters, for instance the
length and the period of observed time series).

Despite the wide variety of empirical evidence of the experience curves, there is a lack of theoretical and mathematical framework of the concept.  Motivated
by this fact and by the fact that there exists a large number of cases where the distribution describing a complex phenomenon is not Gaussian, e.g.~the price fluctuations of most financial assets \cite{Bouchaud}, in this paper we present a theoretical, mathematical framework for describing a probability distribution
function of the volatility of the cumulative production for an {\em arbitrary} probability distribution of noise. In analogy to the
concept of learning curves, which is a relation between the input and the
output of a learning process, one of our main findings shows the relation
between previous and next probability distribution functions which characterizes their volatility.

Knowing distribution functions of the cumulative production and its volatility allows us to understand the
complex behaviour of the system and to calculate the various quantities, such
as mean, variance and also higher order moments, price volatility correlation, etc.

\section*{Volatility for narrow distributions}
It was first discovered by Sahal \cite{Sahal} that the exponentially increasing cumulative production and exponentially decreasing costs gives an experience curve law, which indicates a linear relationship  between cost and increment of cumulative production.

Similar to \cite{Lafond}, let us consider that empirically cumulative production growth follows a smooth exponential behaviour in the presence of noise, by assuming that production is a geometric random walk with drift $g$ and variance $\sigma_a^2$. Within this model, cumulative production is given by: 
\begin{equation}
Z{_t}=\sum_{j=0}^{t}{ {\rm e}^{gj}{{\rm e}^{a_{1}}...{\rm e}^{a_{j}}}},
\end{equation}
where $a_{1}, a_{2}$, ... are stochastic i.i.d.~variables, which describe the
presence of noise in the production process.

Let us first consider the special case, where $a_{1}, a_{2}$, ...~are normally
distributed i.i.d.~variables, with mean zero and variance
$\sigma_a^2$. For the calculation of cumulative production and its volatility in \cite{Lafond} the saddle point method was used.
The main idea of the saddle point is to approximate an integral by
taking into account only the range of the integration where the integrand
takes its maximum. A priory, this can only be correct for small variance
$\sigma_a^2$.

In \cite{Lafond} first the expectation value of cumulative production and its
variance were calculated, which lead to the multiple integral
over $a_i$
\begin{eqnarray}
\label{1}
E(\log Z)&=\int_{-\infty}^{\infty} \log Z\prod_{i=1}^t
\frac{da_i}{\sqrt{2\pi \sigma_a^2}}
\exp\Big[-\frac{a_i^2}{2\sigma_a^2}\Big]\nonumber\\
&=
\int_{-\infty}^{\infty}\prod_{i=1}^t
\frac{da_i}{\sqrt{2\pi \sigma_a^2}}{\rm e}^{S(\{a_i\})},\label{DefProbl}
\end{eqnarray}
with $S(\{a_i\})=\log (\log Z)-\sum_{i=1}^t \frac{a_i^2}{2\sigma_a^2}$.

For $\sigma_a^2\ll 1$ the saddle point method yields explicit results, for
instance the variance of $\log Z(t)$
\begin{eqnarray}
\mbox{Var}(\log Z(t))&=E(\log^2 Z)-E(\log Z)^2\nonumber\\
&=\sigma_a^2
\left(\frac{2 {\rm e}^{g }+1}{1-{\rm e}^{2 g }}+t\right)
+O(\sigma_a^4).
\end{eqnarray}
Finally, the main result of this method is {\em volatility}, i.e.~the variance of {\em volatility
variable} $\Delta \log Z:= \log Z_{t}-\log Z_{t-1}$ for large time $t\rightarrow
\infty $ and is given by the following expression (valid for $g>0$ and small $\sigma_{a}^2$):
\begin{eqnarray}
\mbox{Var}(\Delta \log Z)&=\sigma_a^2 \tanh
\left(\frac{g}{2}\right)+O(\sigma_a^4).
\end{eqnarray}
For details of the derivation we refer to \cite{Lafond}.
Since $\tanh \left(\frac{g }{2}\right)<1$ we have always $\mbox{Var}(\Delta
\log Z)<\sigma_a^2$ an inequality, which means the volatility of cumulative production is lower than the volatility of production.
We have tested this remarkably simple, but potentially powerful relationship using empirically available data and we found that it works reasonably well.

\section*{Volatility for the general case}
The core result of this paper is the investigation of the volatility of
cumulative production for more general distribution functions of $a_{i}$ than
considered in \cite{Lafond}.  Let us assume that in Eq.~($1$), $a_1$, $a_2$,
... are stochastic i.i.d.~variables which are distributed according to some
distribution function $\rho_a$ of any shape and width.

We are interested in the distribution function of the cumulative production
$z_t:=\log Z_t$ which we call $\rho_{z_t}$. From this distribution function we
can calculate all important characteristic quantities of the
system. Surprisingly, the distribution function can be shown
to satisfy a useful recursion relation for successive times:
\begin{multline}
\rho_{z_{t+1}}(x)=\frac 1{1-\exp(-x)}.\\
\rho_{a}*\rho_{z_t}(\log(\exp(x)-1)-g).
\label{wichtig}
\end{multline}

The detailed derivation can be found in Appendix A.

Generally Eq.~(\ref{wichtig}) has to be solved numerically by recursions. Numerical analyses can be done to high accuracy and completely replace simulations, which are time consuming and sometimes inaccurate.

 It is possible to obtain analytic results at least for two cases. In the case of a distribution of $\rho_{a}$ with main weight around some $x_0$ and a
value of $g$ such that $g+x_0>0$ we find for large $t$ an asymptotic
solution. In this case only large values of $x$ matter and (\ref{wichtig}) linearizes to
$\rho_{z_{t+1}}(x)=\rho_{a}*\rho_{z_t}(x-g)$. The second case is a narrow distribution of $\rho_{a}$, which we comment later.

Eq.~(\ref{wichtig}) is highly useful in numerical calculations, especially because the
convolution integral can be carried out efficiently and the
convergence for increasing time is fast. Of course it would be desirable to
treat the time evolution of the probability distribution function for
arbitrary $t$ fully analyticaly such as in \cite{Zadourian}. The
analytical solution is the subject of current investigation.

\begin{figure}
\begin{centering}
%\centerline{\includegraphics[scale=0.6]{bar}}
\hspace*{-0.7cm}  
\centerline{\includegraphics[scale=0.23]{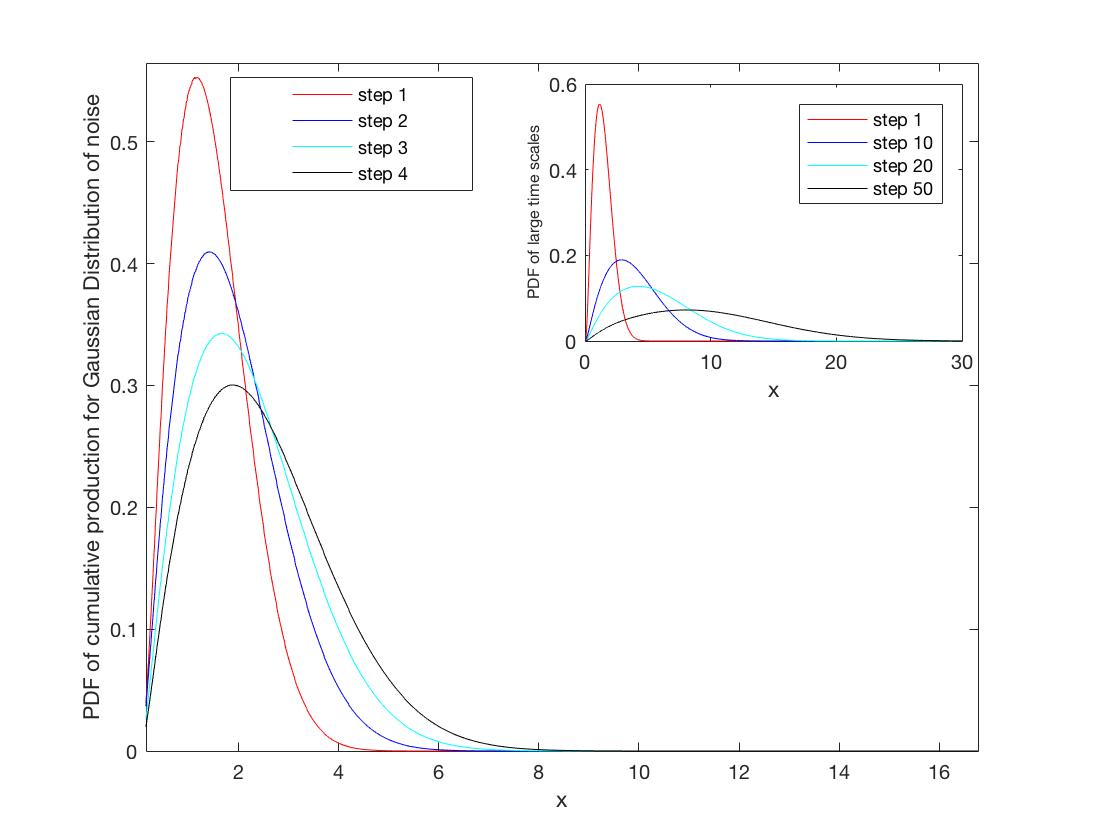}}
\end{centering}
\caption{Plot of probability distribution functions of $\log Z_t$ of the Gaussian distributed $\rho_a$ for different time steps, where $g=0.2$ and $\sigma_{a}=1$ are chosen. The inset shows curves on a large time scale.}
\label{fig1.fig}
\end{figure}

\begin{figure}
\begin{centering}
%\centerline{\includegraphics[scale=0.6]{bar}}
\hspace*{-0.7cm}  
\centerline{\includegraphics[scale=0.23]{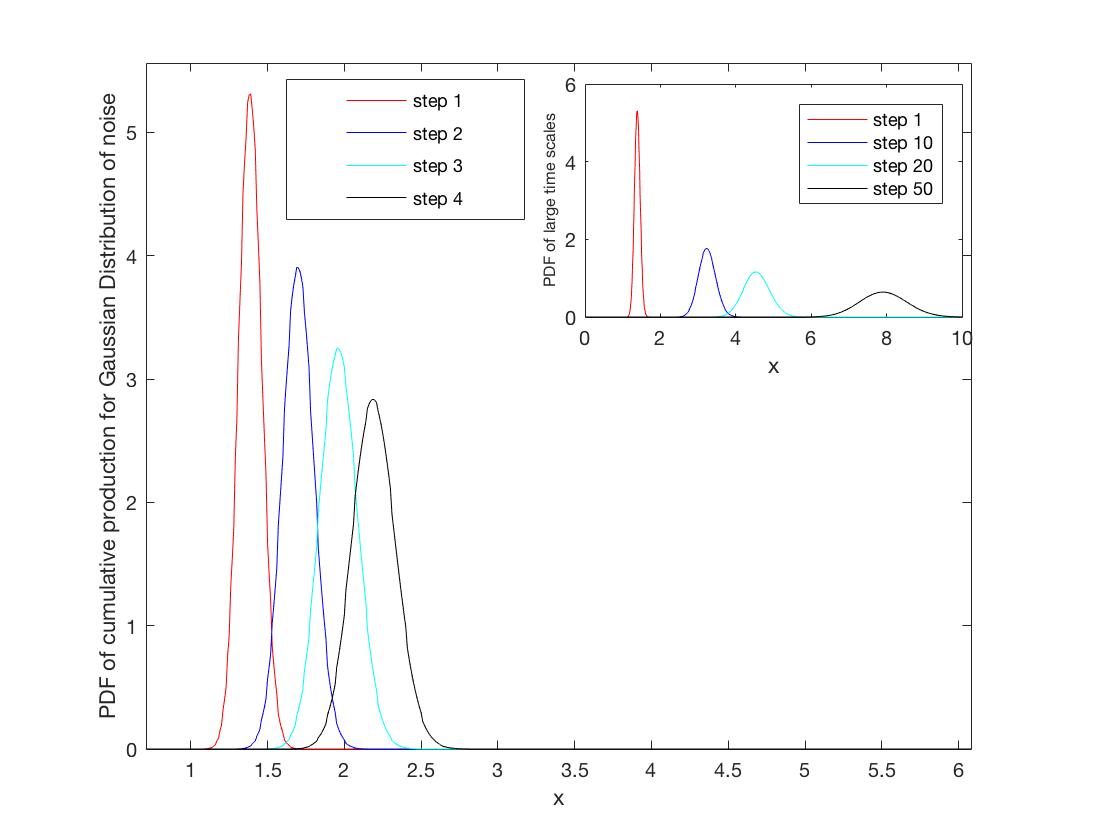}}
\end{centering}
\caption{Plot of probability distribution functions of $\log Z_t$ of Gaussian distributed noise for different time steps, where $g=0.2$ and $\sigma_{a}=0.1$ are chosen. The inset shows curves on a large time scale.}
\label{fig2.fig}
\end{figure}

\begin{figure}
\begin{centering}
%\centerline{\includegraphics[scale=0.6]{bar}}
\hspace*{-0.7cm}  
\centerline{\includegraphics[scale=0.23]{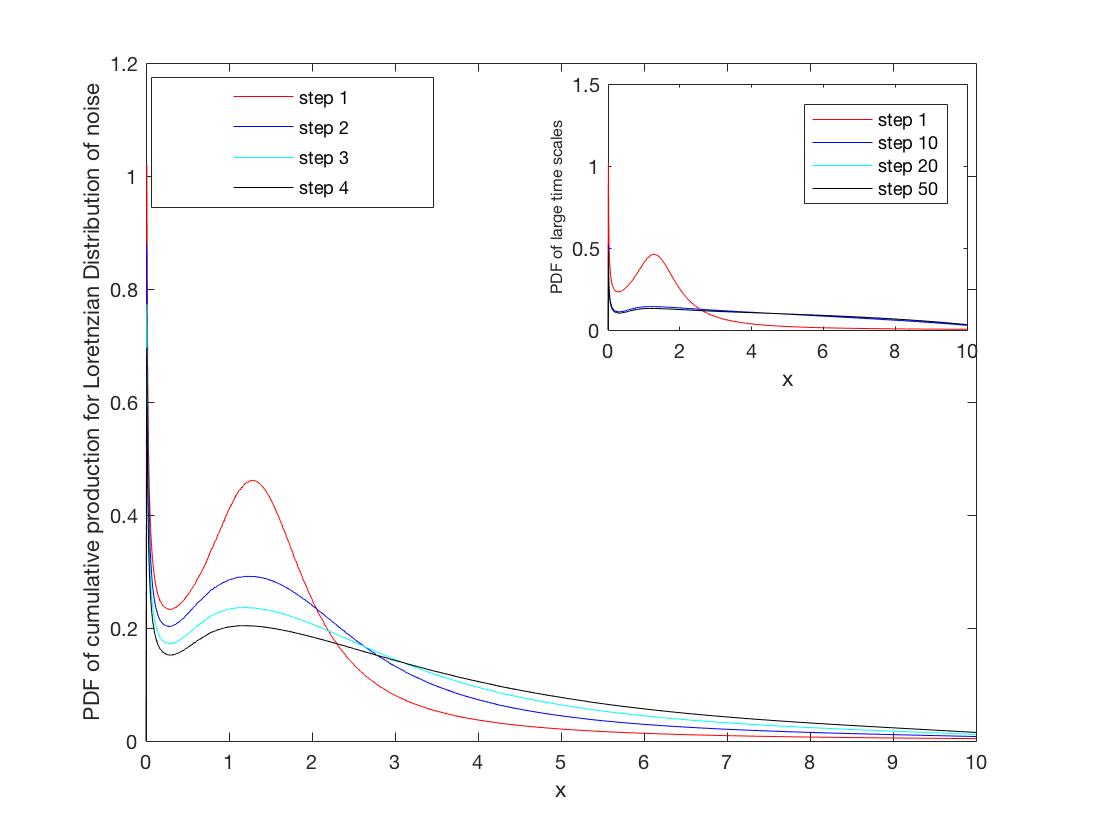}}
\end{centering}
\caption{Plot of probability distribution functions of $\log Z_t$ of Lorentzian distributed noise for different time steps, where $g=0.2$ and width equal to 1 are chosen. The inset shows curves on a large time scale.}
\label{fig3.fig}
\end{figure}

\begin{figure}
\begin{centering}
%\centerline{\includegraphics[scale=0.6]{bar}}
\hspace*{-0.7cm}  
\centerline{\includegraphics[scale=0.23]{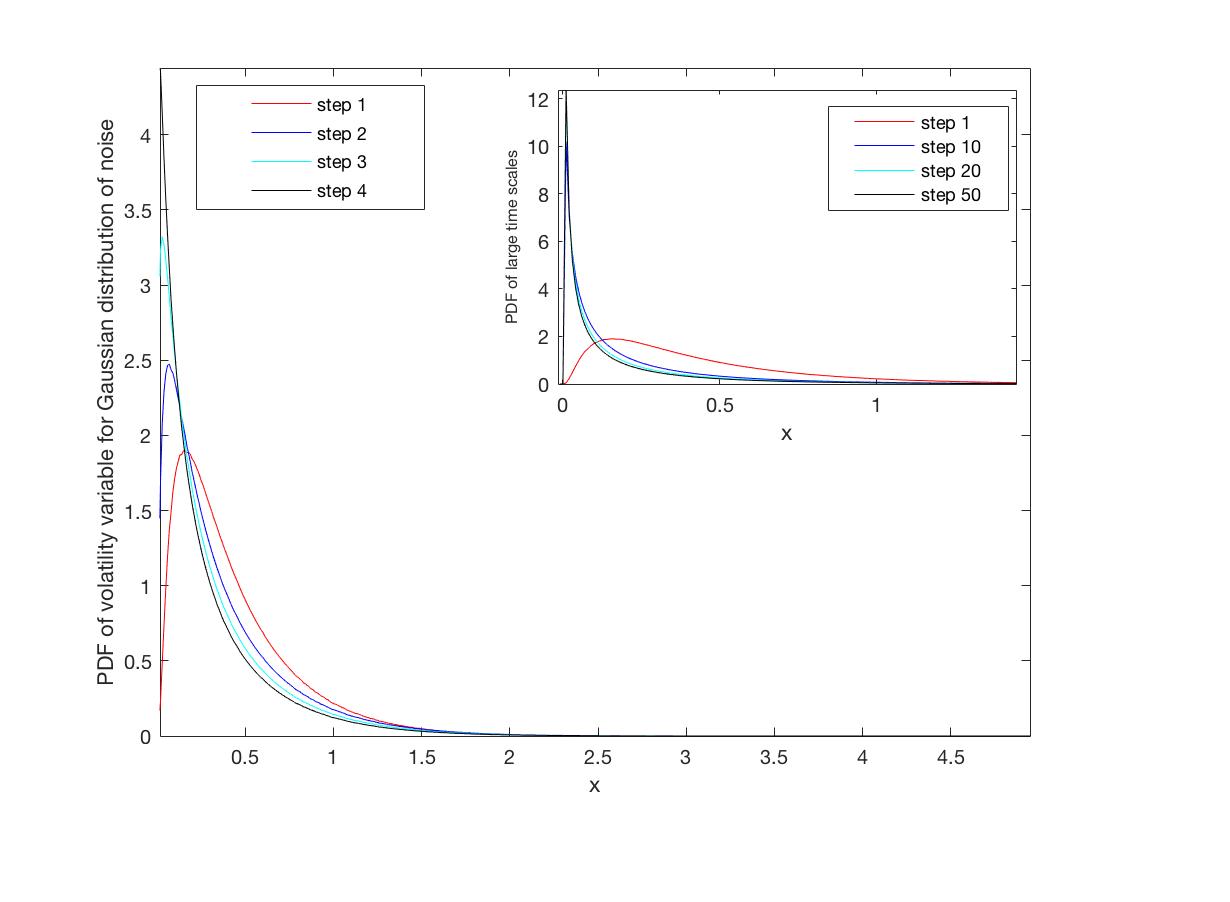}}
\end{centering}
\caption{Plot of probability distribution functions of the volatility variable
  for the Gaussian case for $\rho_{a}$ and different time steps, where $g=0.2$
  and $\sigma_{a}=1$. The Inset is considered for a large time scale.}
\label{fig4.fig}
\end{figure}
\begin{figure}
\begin{centering}
%\centerline{\includegraphics[scale=0.6]{bar}}
\hspace*{-0.7cm}  
\centerline{\includegraphics[scale=0.23]{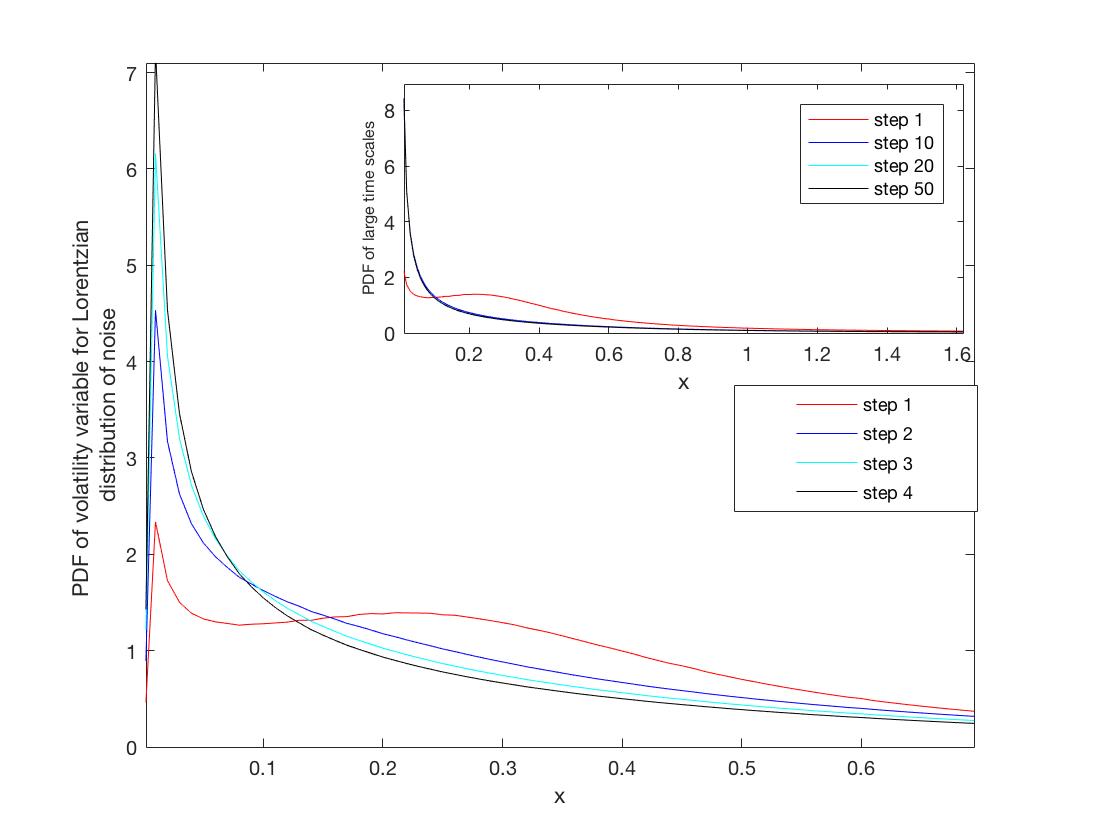}}
\end{centering}
\caption{Plot of probability distribution functions of the volatility variable
  for the Lorentzian case for $\rho_{a}$ and different time steps, where $g=0.2$
  and width is equal to 1. The inset shows curves on a large time scale.}
\label{fig5.fig}
\end{figure}

Fig.~$1$, $2$ and $3$ show the distributions of cumulative production for different types of noise and several time steps. 

\section*{The distribution of volatility} 
The quantity of our interest is the volatility, which is by definition the variance of the distribution function $z_{t}-z_{t-1}$.  We observe
\begin{eqnarray}
\Delta{z_{t}}&:=&z_t-z_{t-1}=\log Z_t-\log Z_{t-1}\nonumber\\
&=&-\log\left(1-\frac{ Z_t-Z_{t-1}}{Z_t}\right).
\end{eqnarray}
We define:
\begin{equation}
Y_{t}:=\frac{Z_t}{Z_t-Z_{t-1}}.
\end{equation}
In Appendix B we show $Y_{t}$ is distributed as $Z_{t}$ in Eq.~($1$) with
$g\to -g$ and $a_{i}\to -a_{i} $, see Eq.~(22).
 
From the above expression we get the following result for the distribution
function of the variable $y_{t}=\log{Y_t}$:
%\rho_{\Delta{z_{t}}}(x)=\frac{1}{x}\rho_{z_{t}^{v}}(-\log{x}),
\begin{equation}
\rho_{\Delta{z_t}}(x)=\frac{1}{\exp(x)-1}\rho_{y_{t}}(-\log(1-\exp(-x))),
\label{eqnine}
\end{equation}
where $\rho_{y_{t}}$ satisfies the same recursion as $\rho_{z_{t}}$, after
changing the signs of $g$ and $a_{i}$, as mentioned above.

The detailed explanation of Eq.~$(\ref{eqnine})$ can be found in Appendix B.

Figs.~$4$ and $5$ illustrate volatility distributions  $\rho_{\Delta{z_t}}$,  considering for
$\rho_{a}$ normal and lorentzian distributions. The Figures show the different behaviours of $\rho_{z_t}(x)$,
with singular (but integrable) characteristics at $x=0$ for
i.e.~$g\geq0$.  The first few time steps show sizable changes whereas only small changes happen at larger times.

In order to demonstrate the usefulness of Eq.~(\ref{wichtig}) and (8), let us derive
from it Eq.~(4), which is the result of the saddle point approximation: we
take for $\rho_a$ a narrow distribution with $0$ mean and (small) variance
${\sigma_{a}^2}$ and accordingly ${\bar y_t}=\log(\sum_{j=0}^t{\rm e}^{-jg})$ and
${\sigma_{t}^2}$ are the mean and the (to be calculated) variance of the
narrow $\rho_{y_{t}}$ distribution. Using variable transformation and
the nice property of additivity of the variance under convolution
we obtain the derivation of the volatility for the special case in Eq.~(4).
We defer the detailed derivation of the above statement to Appendix C.

We also compared our numerical results for general values of $\sigma_a^2$ with
the saddle point result obtained in Eq.~$(4)$. These analyses show that the
general treatment of the probability distributions and the saddle point
approximation coincide for small variance $\sigma_a^2$.

\begin{figure}
\begin{centering}
%\centerline{\includegraphics[scale=0.6]{bar}}
\hspace*{-0.7cm}  
\centerline{\includegraphics[scale=0.23]{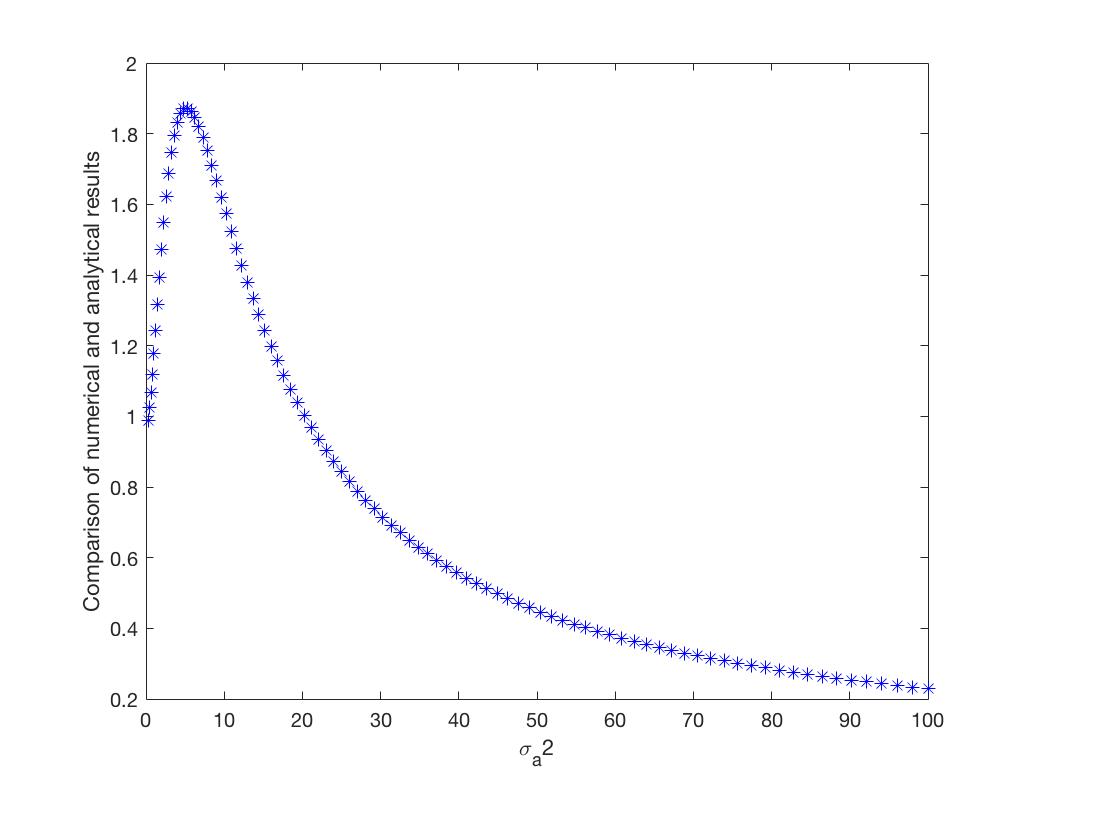}}
\end{centering}
\caption{Plot of the ratio of the numerically exact value of volatility and the saddle point approximation versus various values of the variance ${\sigma_{a}}^2$ of noise, by fixed drift, $g=0.1$.}
\label{fig5.fig}
\end{figure}

Fig.6 shows that for small values of $\sigma_a$ the analytical approximation
and the numerical results coincide whereas for larger $\sigma_a$ we see
sizable deviations. The volatility for intermediate (large) values of
$\sigma_a$ takes larger (smaller) values than the result of the saddle point
approximation.

\section*{summary}
The concept of experience curves has been widely used in different domains, such
as industrial engineering and operations management services, aimed to
estimate the future costs, to reduce the production costs, to evaluate
workers’ learning profile, etc. It plays also a major role in some strategic
tasks, related to capacity, pricing and employment.

Furthermore there exists a large number of cases where the distribution describing a complex phenomenon is not Gaussian. For this variety of applications we found a comprehensive analytical approach, which is based on the probability distribution function of the model. We derived a recursion relation of integral type that replaces simulations by highly accurate numerical integration. The results show how different types of noise affect the cumulative production within the model, based on the learning by doing hypothesis. The distribution functions of the volatility are hardly characterized by the mean and variance and show rather interesting, sometimes singular behaviour.

Knowing such an important quantity fosters a deeper understanding of the
industrial activities. It also allows us to understand the volatile and
complex feature of the system and accordingly to calculate the significant
quantities, by envisioning the opportunities of the model for future
investigations in risk management.

\begin{center}
\text{* * *}
\end{center}

RZ thanks J. Doyne Farmer and Francois Lafond for support and discussions.
\section*{Appendix A: Derivation of the distribution function $\Delta{\rho_{z_t}}$}
Here we will derive the recursion expression (\ref{wichtig}).  Note that the modified
object
\begin{equation}
\tilde Z_t:=\sum_{j=0}^t {\rm e}^{g j}{\rm e}^{a_2}...{\rm
e}^{a_{j+1}}
\end{equation}
has the same distribution function as $Z_t$, because we have used different,
but independent and identically distributed $a_i$'s. So $\tilde z_t:=\log
\tilde Z_t$ is distributed according to $\rho_{\tilde z_t}=\rho_{z_t}$.  Note
that
\begin{multline}
1+{\rm e}^{g+a_1}\tilde Z_t=
1+\sum_{j=0}^{t} {\rm e}^{g (j+1)}{\rm
e}^{a_1}{\rm e}^{a_2}...{\rm e}^{a_{j+1}}\\
=1+\sum_{i=1}^{t+1} {\rm e}^{g i}{\rm
e}^{a_1}{\rm e}^{a_2}...{\rm e}^{a_{i}}
=Z_{t+1}.
\end{multline}
Therefore we have
\begin{equation}
z_{t+1}=\log(1+\exp(g+a_1+\tilde z_t)=f(g+a_1+\tilde z_t)
\end{equation}
where we have used the definition of the function $f$:
\begin{equation}
f(x):=\log(1+\exp(x)).
\end{equation}
The stochastic variable $a_1+\tilde z_t$ is distributed according to the
convolution of $\rho_a$ with $\rho_{z_t}$. The distribution of $g+a_1+\tilde
z_t$ is the convolution with a subsequent shift of the argument:
\begin{eqnarray}
&\rho_{a_1+\tilde z_t }=\rho_{a_1}*\rho_{\tilde z_t}=\rho_{a}*\rho_{z_t},\nonumber\\
&\rho_{g+a_1+\tilde z_t }(x)=\rho_{a}*\rho_{z_t}(x-g).
\end{eqnarray}
With (11) and (13) we can calculate the distribution function $\rho_{z_{t+1}}$
of $z_{t+1}$. If we use the arguments $x$ for $\rho_{g+a_1+\tilde z_t }$ and
$y=f(x)$ for $\rho_{z_{t+1}}$ we find
\begin{equation}
\rho_{z_{t+1}}(y)dy = 
\rho_{g+a_1+\tilde z_t }(x)dx,
\end{equation}
and from this
\begin{multline}
\rho_{z_{t+1}}(y)=[f'(x)]^{-1}\rho_{g+a_1+\tilde z_t }(x)\\
=[f'(x)]^{-1} \rho_{a}*\rho_{z_t}(x-g).
\end{multline}
Now we use $f'(x)=\exp(x)/(1+\exp(x))$ and $x=f^{-1}(y)=\log(\exp(y)-1)$ and
reach one of our main findings:
\begin{multline}
\rho_{z_{t+1}}(x)=\frac 1{1-\exp(-x)}.\\
\rho_{a}*\rho_{z_t}(\log(\exp(x)-1)-g).
\end{multline}

%\noindent

%{\bf Derivation of the distribution function of volatility $\Delta z_t$}\\
\section*{Appendix B: Derivation of the distribution function of the volatility variable $\Delta z_t$}
According to the definition of $Y_t$ we have $Y_t=\frac{Z_{t}}{Q_{t}}$, with
\begin{equation}
Z_t:=\sum_{j=0}^t Q_j,\qquad Q_j:={\rm e}^{gj}{\rm e}^{a_1}...{\rm e}^{a_j}
\end{equation}
where $g$ is a constant, and $a_1$, $a_2$, ... are stochastic i.i.d.~variables
which are distributed according to some distribution function $\rho_a$.  Now
-- luckily -- $Y_{t}$ has the same structure as $Z_t$ if we replace the constant
$g$ and the stochastic variables $a_i$ by $g$ and $-a_i$:
\begin{equation}
Y_{t}:=\frac{Z_t}{Q_t}=\sum_{j=0}^t \frac{Q_j}{Q_t},\qquad 
\frac{Q_j}{Q_t}={\rm e}^{g(j-t)}{\rm e}^{-a_{j+1}}...{\rm e}^{-a_t}
\end{equation}
Next we define
\begin{multline}
\tilde{g}:=-g,\qquad \tilde a_1:=-a_t,\quad \tilde a_2:=-a_{t-1},\quad ...,\\
\tilde a_t:=-a_{1}
\end{multline}
And indeed
\begin{equation}
\frac{Q_j}{Q_t}={\rm e}^{g (j-t)}{\rm e}^{-a_{j+1}}...{\rm e}^{-a_t}
={\rm e}^{\tilde{g} (t-j)}{\rm e}^{\tilde a_1}...{\rm e}^{\tilde a_{t-j}}
\end{equation}
Hence
\begin{multline}
Y_t=\sum_{j=0}^t{\rm e}^{\tilde{g} (t-j)}{\rm e}^{\tilde
a_1}...{\rm e}^{\tilde a_{t-j}}=\sum_{j=0}^t{\rm e}^{\tilde{g} j}{\rm
e}^{\tilde a_1}...{\rm e}^{\tilde a_{j}}
\end{multline}
As mentioned above we find that the distribution function of $y_t:=\log{Y_t}$
corresponds to that of $z_t:=\log{Z_t}$, by taking into account sign changes
of $g$ and $a_i$. Hence it satisfies the recursion relation derived in App.~A:
\begin{multline}
\rho_{y_{t+1}}(x)=\frac 1{1-\exp(-x)}.\\
\rho_{-a}*\rho_{y_t}(\log(\exp(x)-1)+g).
\end{multline}

%\section*{Derivation of the volatility for the narrow distribution from the
%  general formalism}
\section*{Appendix C: Analytic treatment of narrow distributions}
For obtaining the saddle point formula Eq.~(4) from the general recursion
relation we need to recall Eqs.~(5) and (22).  $\rho_a$ is given by a narrow
distribution around $0$ and variance ${\sigma_a}^2$ and correspondigly
$\rho_{y_t}$ is defined by the mean ${\bar y_t}=\log(\sum_{j=0}^t{\rm
  e}^{-jg})$ and ${\sigma_t}^2$.

Due to the additivity feature of the variance under convolution, the narrow
$\rho_{a}*\rho_{y_t}$ has the
variance equal to ${\sigma_t}^2+{\sigma_a}^2$.

Let us use the variable transformation in Eq.~(22). Then we get  ($\sigma_t:=\sigma_{y_t}$)

\begin{equation}
\frac{d(\log({\rm e}^x-1)+g)}{dx}{\sigma_{t+1}}=\sqrt{{\sigma_t}^2+{\sigma_a}^2},
\end{equation}
where $x=\bar y_t$. For large times $t \to \infty$
we get 
\begin{equation}
{\sigma_\infty}^2=\frac{1}{{\rm e}^{2g}-1}{\sigma_a}^2.
\end{equation}
Let us now calculate the main result, namely $\sigma_{\Delta{z_{t}}}$, the
volatility for narrow distributions. To this end we need to transform
variables in Eq.~(5)
\begin{equation}
\sigma_{\Delta{z_{t}}}=({\rm e}^x-1){\sigma_t}
\end{equation}

For $x$ equal to its maximum we have:
\begin{equation}
-\log(1-{\rm e}^{-x})=\bar y_{t}
\end{equation}
and for $t \to \infty $ this amounts to $x=g$.

Finally, we obtain the result for the special, narrow distributed function,
which in the previous work \cite{Lafond} was obtained by the saddle point
approximation.
\begin{equation}
\sigma_{\Delta{z_{t}}}=\sqrt{\tanh{(\frac{g}{2}})}{\sigma_a}.
\end{equation}

%$\frac{\sigma{conv}}{\sigma_{t+1}}=\frac{dy}{dx}$

\end{document}